\documentclass
[11pt]{article}
\usepackage{tipa}
\usepackage{amssymb}
\usepackage{mathrsfs}
\usepackage{amsfonts}
\usepackage{amsmath}
\usepackage[top=1in, bottom=1in, left=1.25in, right=1.25in]{geometry}
\usepackage{graphicx}
\begin{document}

\title{\textbf{Network Evolution by Relevance and Importance Preferential Attachment}}
\author{Weituo Zhang, Chjan Lim}

\maketitle
\begin{abstract}
Relevance and importance are the main factors when humans build
network connections. We propose an evolutionary network model based
on preferential attachment(PA) considering these factors. We analyze
and compute several important features of the network class
generated by this algorithm including scale free degree
distribution, high clustering coefficient, small world property and
core-periphery structure. We then compare this model with other
network models and empirical data such as inter-city road
transportation and air traffic networks.
\end{abstract}

\section{Introduction}

For many complex networks in society, it is arguable that
the \emph{Relevance} and the \emph{Importance} are the two main factors
influencing how new network connections are formed in existing
dynamic networks. One typical scenario is in scientific research and
the  publication process. In choosing references, authors are more
likely to cite articles with high impact(importance) and also those
using similar method or discussing relevant issues(relevance).
Another example is in the design and organic growth of intercity
transportation networks. Traffic engineers and city designers prefer
to connect a given city to big cities with high
connectivity(importance) but also want to reduce the expense by
giving priority to the connections between nearby cities(relevance).
Complex networks involving both relevance and importance also
include aspects of the World Wide Web(WWW) and many social networks.
An interesting and ironic point is that people are still striving to
understand the properties of these complex networks which are
largely man-made. As a related point, we emphasize that the network
evolutions studied here are governed by a distributed decision-
making system rather than centrally organized. For each agent in the
networks that makes local decisions, the rule of adding or deleting
links may be simple and clear. Intuitively, the complexity of the
networks arise from some other reasons such as cooperative and bulk
properties of large systems consisting of many similar subunits.
While this complexity is not explicit in the local design rules and
is often beyond total control of the network designers, human
society nonetheless seek to understand and manage this complexity.
Hence, the current scientific and technological interests in
studying the origins and properties of these dynamic complex
systems. In this paper, we suggest that one of the origins of this
complexity is an underlying metric space defining the
\emph{Relevance} structure which we will introduce and discuss in
detail later.

In the past few decades, several evolutionary network models have
been proposed with respect to one or both of the two factors,
\emph{importance} and \emph{relevance}. For \emph{importance} alone,
the most famous model is invented by Barabasi, Albert, known as BA
network model\cite{BA} or ``preferential attachment''(PA) algorithm.
The standard preferential attachment starts with a network with
$N_0$ vertices and $m_0$ edges. New vertex is successively added and
attached to $m<N_0$ preexisting vertices. The probability of
attaching to a vertex $i$ is proportional to its degree $k_i$. This
algorithm will naturally generate the network with power-law degree
distribution $p(k)\sim k^{-\gamma}$ with $\gamma=3$. There are many
variations of the PA algorithm in the
literature\cite{Dangalchev,Fortunato,Alejandro}, all of which have
the relevant complexity. From this, we conclude that the
preferential attachment to high degree nodes, i.e. the ``rich get
richer'' effect, is the essential reason for the emergence of scale
free degree distribution.

There are also well known network models based on the \emph{relevance}.
The simplest evolutionary model considering the
\emph{relevance} is the Random Geometric Graph(RGG). In this model,
we successively add vertices at random locations in a unit square,
and link each new vertex to all the nearby vertices within a given
radius $r$. Here the \emph{relevance} is measured by the geometric
distance. Another model considering the \emph{relevance} is given in
\cite{Community}, in which the \emph{relevance} is given by a
hierarchical structure and tree distance.
According to these models, we find the most natural way to measure the \emph{relevance} is to suggest an underlying metric space. We will show later in this paper that how this metric space affects the global properties of the network. Due to the triangle inequality in metric spaces, the corresponding relevance relationship satisfies that any two objects relevant to the same one should also be relevant each other. Therefore, network models based on the \emph{relevance} often have high clustering coefficients. The other way of thinking the relevance structure of the complex
networks is to use geometric
embedding\cite{Daniel,Janssen,Hyperbolic} which is not to provide an
evolutionary model but to find the most suitable underlying metric
space for the known network.

Motivated by these network models, we propose an evolutionary network model with appealing properties that takes the both two factors into consideration. Our model is based on the preferential attachment. But beside the preferential attachment to the high degree nodes, we also suggest the preferential attachment to nearby nodes under a given metric space. Later in this paper, we will introduce our Relevance and Importance Preferential Attachment(RIPA) model given by an evolution process, analyze several network properties, and compare this model with other network models and some empirical data.

\section{Model}
In this section we will describe the algorithm called Relevance and Importance Preferential Attachment(RIPA) which generates a class of complex networks. The RIPA, similar to the classical preferential attachment, starts with a initial network with $N_0$ vertices and $m_0$ edges. A new vertex is attached to $m$ other vertices with the probability depending on the importance and relevance of those vertices.

In RIPA, the importance of a vertex is valued by its degree as in the classical preferential attachment, and the relevance is given by a metric space $\Omega$. We denote distance between two elements $x,y\in \Omega$ by $d(x,y)$. Then $\rho(x,y)$, the relevance between them, is defined as a non-increasing function of the distance $d(x,y)$,
$$\rho(x,y)=f(d(x,y)),$$

satisfying $f(0)=1$ and $f(\infty)=0$. A typical example is $f(x)=e^{-x}$, but $f$ can also have a power-law tail.The centrality defined below measures the general influence of an element $x$ on the whole space.
\begin{equation}
C(x)=\int_\Omega \rho(x,x')dx'.
\label{centrality}
\end{equation}

Centrality actually gives, in another sense, an ``importance'' according to the position in the underlining metric space instead of the connectivity to other vertices. In the scenario of the between-city transportation, centrality measures the physical geographical transportation condition of a position. In the scenario of scientific research, a research topic has high centrality means it is a bridge of many other fields and therefore is important by itself regardless how it is recognized by citations. Later in this paper, we will investigate some cases on the metric spaces with constant centrality $C(x)\equiv C$. Examples are: (1)square with periodic boundary condition, (2) sphere in 3-d space
In these spaces, there is no ``center'' position and every element is at an equivalent place.

A further restriction here for the relevance $\rho$ and hence $f$ is that the integral in Eqn.~(\ref{centrality}) should be well-defined. This restriction is fairly important especially when we consider the large network limit.

In the RIPA, a new vertex $j$ is attached to the preexisting vertex $i$ by the probability
$$\Pi_{ij}={k_i\rho_{ij}\over z(x_j)}.$$
Here $k_i$ is the degree of $i$ indicating the importance and $\rho_{ij}=\rho(x_i,x_j)$ is the relevance between $i,j$. $z(x_j)$ is the normalization constant so that $\sum_i \Pi_{ij}=1$. $z(x)$ is defined as a function on $\Omega$ called local partition by
$$z(x)=\sum_{i} k_i \rho(x_i,x).$$
The summation here goes over all existing vertices. A particular position $x\in\Omega$ with higher local partition $z(x)$ has more overall relevance to previous vertices, therefore may attract more interest of a new vertex. So we suggest $\mu(x)$, the probability of emergence of a new vertex at $x$, is proportional to $z(x)$,
$$\mu(x)={z(x)\over Z},$$
where $Z$ is the global partition function
$$Z=\int_\Omega z(x) dx=\int_\Omega\sum_j k_j \rho(x_j,x) dx=\sum_j k_jC_j.$$
In a metric space with constant centrality, we further have $Z=K C$ where $K=\sum_i k_i=m_0+mt$ is the total number of degree in the network and grows linearly with time $t$.

We summarize the algorithm of RIPA as follows:
\begin{itemize}
\item 1. Begin with a network with $N_0$ nodes.
\item 2. For $j=N_0+1$ to $N$
\subitem 2.1 Add a new node $j$ at the position $x$ with probability $\mu(x)={z(x)\over Z}.$
\subitem 2.2 Attach $j$ to $m$ preexisting nodes $i$ with probability $\Pi_{ij}={k_i\rho_{ij}\over z(x_j)}.$
\end{itemize}

The expected change of the degree of the vertex $i$ is given by
$$E\left[{dk_i\over dt}\right]=\int_{\Omega}\Pi_{ij}m\mu(x_j)dx_j=\int_\Omega m{k_i  \rho_{ij}\over z(x_j)}{z(x_j)\over Z}dx_j= m{k_i C_i\over Z}.$$
The above equation shows that the degree of a vertex grows at a expected speed proportional to the current degree which is exactly the relation we have in standard preferential attachment algorithm. Therefore we derive the stable degree distribution using the same approach for standard PA and obtain the following theorem:\\

\textbf{Theorem:} \emph{With arbitrary metric space and initial network, the network generated by RIPA has the power-law degree distribution $p(k)\sim k^{-\gamma}$ with $\gamma=3$ as $t\rightarrow \infty$.}\\

Besides, the change of the local partition $z(x)$ comes from two parts: the growth of degrees of the existing vertices and the new vertex. When the centrality is constant $C$, we have
\begin{eqnarray*}
E\left[{\partial z(x)\over \partial t}\right] &=& \sum_i E\left[{dk_i\over dt}\right] \rho(x_i,x) + m\int_\Omega \rho(x',x) \mu(x')dx'\\
                                              &=& {C \over Z} \left(z(x)+m \bar{z}(x)\right).
\end{eqnarray*}
Here $\bar{z}(x)={1\over C}\int_\Omega z(x')\rho(x',x)dx'$ is considered as an average of $z$ in the neighborhood of $x$ by the weight function $\rho(x',x)$. The above equation can be rewritten as
$$E\left[{\partial z(x)\over \partial t}\right]={C \over Z} \left[(m+1)z(x)+m (\bar{z}(x)-z(x))\right].$$
On the right hand side, the first term is respect to exponential growth tending to generate a scale free distribution of $z(x)$, the second term is a diffusion term which will smooth the distribution of $z(x)$.

Several previous models can be more or less considered as special cases of our model. By selecting the trivial metric space, a singleton, our model becomes the standard BA model. On the unit square with the Euclidean metric, taking the relevance function

$$f(x)=\left\{\begin{array}{ll}
1 & x\in [0, r]\\
0 & \mbox{else}
\end{array}\right. ,$$
we obtain the RGG model. On the tree(graph) as a discrete metric space with the standard tree distance, we obtain the Community Guided Attachment(CGA) model\cite{Community}. In this paper, we mainly investigate the case on the 2-D surface with the Euclidean metric and power-law or exponentially decreasing relevance function.

Next we will discuss several important properties of the RIPA network model. Although there is no rigorous definition of complex networks, many people consider the following three are the typical properties of complex networks: power-law degree distribution (scale free), high clustering coefficient (clustering), short average path-length (small world). A lot of efforts have been made to find network models which capture these properties. The following table summarizes the properties of several known network models. (ER stands for Erdos每Renyi model\cite{ER}, BA stands for Barabasi每Albert model\cite{BA}, RGG stands for Random Geometric Graph\cite{RGG}, WS stands for Watts-Strogatz model\cite{WS}, RAN stands for Random Apollonian network\cite{RAN}.)

\begin{table}
\caption{Summary of the properties of network models}
\begin{center}
\begin{tabular}{cccc}
\hline
Network Model & scale free & clustering& small world\\\hline
ER            &&&$\surd$\\
BA            &$\surd$&&$\surd$\\
RGG           &&$\surd$&\\
WS            &&$\surd$&$\surd$\\
RAN           &$\surd$&$\surd$&$\surd$\\
RIPA          &$\surd$&$\surd$&$\surd$\\
\hline
\end{tabular}
\end{center}
\end{table}

Till now, not many network models satisfactorily capture all of the three typical properties. Some network models like Random Apollonian Network(RAN) do, but is totally artificial without revealing the mechanism from which all the properties of the real world networks come. The RIPA model we proposed here have all of the three properties under certain conditions, and at the same time provides a natural reasoning of these properties. Further more, it also has a core-periphery structure which is an important feature of some real world networks like the world airline network (WAN).

\section{Equivalent model}

In this section, we propose an equivalent model of RIPA called Relevance and Importance Branching Process (RIBP). This model is useful in analyzing some important network properties, especially the degree-degree correlation. RIBP is a random process with continuous time variable $\tau$. Each node $i$ undergoes an independent Poisson process $n_i(\tau)$ with rate $\lambda_i=k_iC_i$. $n_i(0)=0$. When $n_i$ has an increment at time $\tau$, the node $i$ will generate a children node $j$ at a random position $x$ which satisfies the distribution $\rho(x,x_i)\over C_i$. Attach $j$ to $i$, and also to other $m-1$ nodes as the way in RIPA. To compare these two models, we set the relationship between the time variables as
$$t(\tau)=\sum_i n_i(\tau).$$
For RIBP, there is a 1-1 map between the network state and the time variable $t$ as given above, therefore this setting is well defined. We only care about the updates of the network state, so we can rewrite the algorithm of RIBP as follows:
\begin{itemize}
\item 1. Begin with a network with $N_0$ nods.
\item 2. For $j=N_0+1$ to $N$
\subitem 2.1 Pick a node $i$ as the generator with probability $k_iC_i\over \sum_i k_iC_i$.
\subitem 2.2 Locate the new node $j$ at the position $x$ with probability $\rho(x,x_i)\over C_i $, and attach $j$ to $i$.
\subitem 2.3 Attach $j$ to $m-1$ preexisting nodes with probability $\Pi_{ij}={k_i\rho_{ij}\over z(x_j)}.$
\end{itemize}

The crucial difference between RIBP and RIPA lies in the arrival of new nodes. For RIPA, the arrival of new nodes follows a global probability distribution which is affected by all the existing nodes and thus is very complicated. For RIBP, the arrival of of new nodes is more like what happened in some private clubs: the membership of a new guest requires the invitation of an existing member and there is a default social link between the new member and his/her inviter. The RIBP is more parallelizable because each existing node invites new nodes to join the network independently and the location of the new node is only affected by its inviter. Therefore, compared to RIPA, the RIBP is much easier to implement and analyze. As we will show below, by carefully choosing the rate of the invitation for each node, we build up the RIBP which is essentially a different stochastic process from RIPA but generate the same random network ensemble.

Let $G(t), G'(t)$ denote the two random processes of evolutionary networks generated by RIPA and RIBP respectively. We define the equivalence of the two models as:

 \emph{If at time $t_0$, $G(t_0)=G'(t_0)$, then $G(t)$ and $G'(t)$ as random networks have the same probability measure at $t>t_0$.}

To prove the equivalence we only need to show that, given $G(t_0)=G'(t_0)$, for any $G_0$, the two probabilities are equal
$$P\left(G(t_0+1)=G_0\right)=P\left(G'(t_0+1)=G_0\right).$$

For simplicity, we first consider the $m=1$ case. Without loosing any generality, let $G_0$ be the state that the next new node $j$ locates at $x_0$ and attaches to the node $i$. Therefore
\begin{eqnarray*}
P\left(G(t_0+1)=G_0\right) &=&\mu(x_0)\Pi_{ij}={z(x_0)\over Z}{k_i \rho(x_i,x_0)\over z(x_0)}={k_i \rho(x_i,x_0)\over Z}.\\
P\left(G'(t_0+1)=G_0\right)&=&{k_i C_i\over \sum_i k_i C_i}{\rho(x_i,x_0)\over C_i}={k_i \rho(x_i,x_0)\over Z}.
\end{eqnarray*}

More generally, when $m\ge 1$, $G_0$ is the state that the next new node $j$ locates at $x_0$ and attaches to the nodes $i_1,i_2,...,i_m$.
\begin{eqnarray*}
P\left(G(t_0+1)=G_0\right) &=&\mu(x_0)m! \Pi_{i_1j}...\Pi_{i_mj}={m!\over Z z^{m-1}(x_0)}{\prod_{i=i_1...i_m} k_i \rho(x_i,x_0)}\\
\end{eqnarray*}
To calculate $P\left(G'(t_0+1)=G_0\right)$, we first consider the case that $i_1$ is the generator. The probability for this case is
$$P_{i_1}={k_{i_1} C_{i_1}\over \sum_i k_i C_i} {\rho(x_0,x_{i_1})\over C_{i_1}} (m-1)! \Pi_{i_2j}...\Pi_{i_mj}={(m-1)!\over Z z^{m-1}(x_0)}{\prod_{i=i_1...i_m} k_i \rho(x_i,x_0)}. $$
Sum up the probabilities of all such cases, we have
$$P\left(G'(t_0+1)=G_0\right)=m P_{i_1}=P\left(G(t_0+1)\right).$$

When $m=1$, it is easy to show that all the degrees $k_i$'s are independent by construction. When $m\ge 2$, the degrees are correlated and then we will analyze the degree-degree correlation at fixed time $t$. \\

At time step $t$, we calculate the probability for the new node attach to node $i$ and $j$ at the same time.
$$P_{actual}=\int {2\over Z z(x_0)}{k_i\rho(x_i,x_0)}{k_j\rho(x_j,x_0)}dx_0={2k_ik_j\over Z}\int{\rho(x_i,x_0)\rho(x_j,x_0)\over z(x_0)} dx_0$$

As a baseline, we also calculate the probability for the same event under the assumption that $k_i(t)$, $k_j(t)$ are independent.
$$P_{indep}=2 \int{k_i\rho(x_i,x_0)\over Z} dx_0 \int{k_j\rho(x_j,x_0)\over Z} dx_0={2k_ik_j\over Z^2}\int{\rho(x_i,x_0)} dx_0\int{\rho(x_j,x_0)} dx_0$$

$\Delta=P_{actual}-P_{indep}$ implies the correlation between $k_i$ and $k_j$. When $\Delta>0$, $k_i$ and $k_j$ are correlated; When $\Delta<0$, $k_i$ and $k_j$ are anti-correlated.
$$\Delta={2k_ik_j\over Z^2}\left[{\int z(x_0) dx_0 \int{\rho(x_i,x_0)\rho(x_j,x_0)\over z(x_0)} dx_0-\int{\rho(x_i,x_0)} dx_0\int{\rho(x_j,x_0)} dx_0}\right].$$

According to the above formula, we conclude that degrees of two nodes can be anti-correlated due to the following two reasons:\\
1.  The two nodes are far apart such that $\rho(x_i,x_0)$ and $\rho(x_j,x_0)$ are anti-correlated, i.e. relevant to one means irrelevant to the other. \\
2.  The two nodes are separated by a high $z(x)$ area, hence the new nodes relevant to the both are very likely to be distracted by the nodes in between.\\

\section{Between-city transportation}
In this section we focus on RIPA on 2-dimensional surface with respect to the case of between-city transportation. First, we consider networks generated by RIPA on the unit square $D$ with periodic boundary conditions. The relevance $\rho$ is given by $f(x)=\exp{(-\lambda x)}$. In this case the total partition function is:
$$Z=\int_{x\in D}\sum_{j=1}^N k_j e^{-\lambda d(x_j,x)} dx$$

\begin{figure}[!htbt]
\caption{Network generated on unit square with periodic boundary condition. $m=1$, $N=5000$, $\lambda=10$. The circles are centered at the locations of the cities and the radii represents their degrees. The background color indicates the logarithm of local partition. }
\vspace{0.5cm}
\begin{center}
\includegraphics[width=0.85\textwidth]{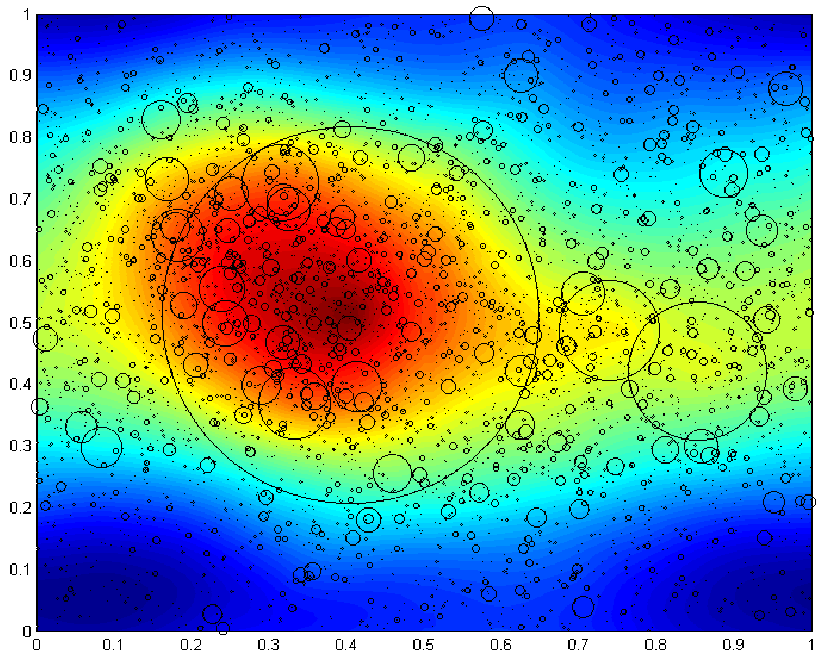} \\
\label{square}
\end{center}
\end{figure}

Figure \ref{square} represents a special realization of the network. Each circle in the figure represents a city, the center of the circle indicates the locations of the city and the radius indicates the degree, the color(brightness) in the background indicates the logarithm of the local partition function $z(x)$. In Fig.\ref{square}, we observe a phenomenon that cities tends to gather but big cities tends to separate. For example, around the greatest city (the capital), we can find bigger city in the area further from the capital. This is because a huge city has two effects: (1) the local partition in its neighbor area is bigger therefore attract more new cities, (2)it will attract more links from new cities therefore inhibit the nearby cities to grow. The second effect is the most significant when we choose small $m$.

Next, we will investigate the properties of the RIPA network model one by one in this special case, and compare this network model with the BA network and the world airline network (WAN)\cite{WAN}. The later is an empirical network from openfights.org.

\subsection{Degree distribution}
\begin{figure}[!htbp]
\caption{Power-law degree distribution of networks when $m=1,5$, $N=5000,10000,20000$, $\lambda=10$. }
\begin{center}
\includegraphics[width=0.85\textwidth]{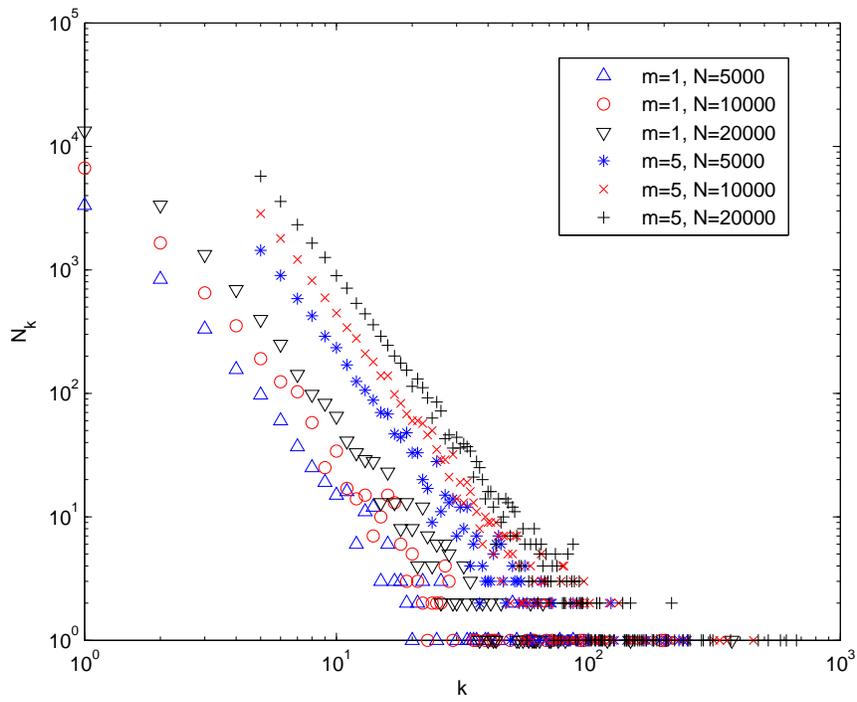} \\
\label{scalefree}
\end{center}
\end{figure}

Fig. \ref{scalefree} shows that the power-law degree distribution of the RIPA network. As analyzed before, the degree distribution is $N_k\sim k^{-\gamma}$. $N_k$ is the number of vertices with the degree $k$. The index $\gamma=3$ as the same as in the BA network model.

\subsection{Clustering Coefficient}
\begin{figure}[!htbp]
\caption{Clustering coefficients $C$ as a function of network size $N$ for different types of networks. RIPA1 with m=3, RIPA2 with m=10.}
\begin{center}
\includegraphics[width=0.85\textwidth]{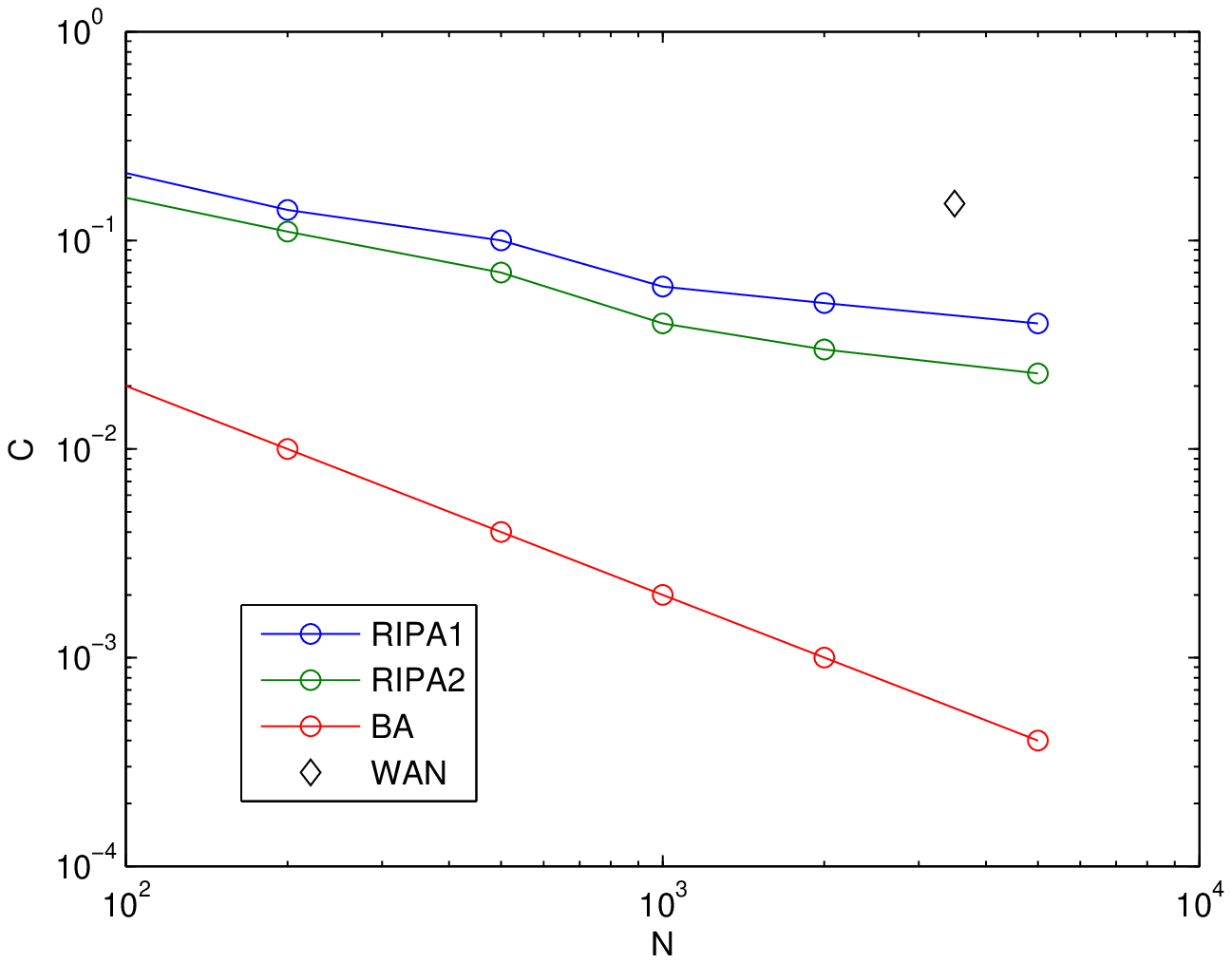} \\
\label{clustering}
\end{center}
\end{figure}

The clustering coefficient quantifies how well connected are the neighbors of a node in a network. Network models considering the \emph{relevance} usually has higher clustering coefficient than purely random networks. This is because the \emph{relevance} is naturally transitive, i.e. two objects relevant to the same thing are more likely to be relevant to each other. Consequently, the RIPA network has a significant higher clustering coefficient then the ER or BA networks. Fig.\ref{clustering} shows the clustering coefficients of the RIPA network, the BA network and the WAN network\cite{WAN}.

\section{Average path-length}
In the area of complex networks, we say a network is a ``small world'' if the average path-length of two arbitrary nodes in the network is no more than the order $O(\ln(N))$ as the network size $N$ grows. There are two different large $N$ limits of this network model. One is the non-extensive limit, for which the metric space keeps the same and the density of nodes increases to infinity. The other is the extensive limit, for which the density of nodes keeps the same and the metric space extends to infinity. In the latter case, an equivalent way is to keep the metric space the same and rescale the metric. For instance, on the unit square, the metric $d(x,y)$ should be rescaled as $d_N(x,y)={\sqrt{N}}d(x,y)$, so that the average density of nodes keeps constant as $N$ grows.

\begin{figure}[!htbp]
\caption{Average path-length $L$ in RIPA network as network size $N$ grows. Red plots are for the RIPA under the non-extensive large $N$ limit. Blue and Green plots are for the RIPA under the extensive large $N$ limit. The blue plot is for the relevance function with power-law decay, the green one is for the relevance function with exponential decay.}
\begin{center}
\includegraphics[width=0.85\textwidth]{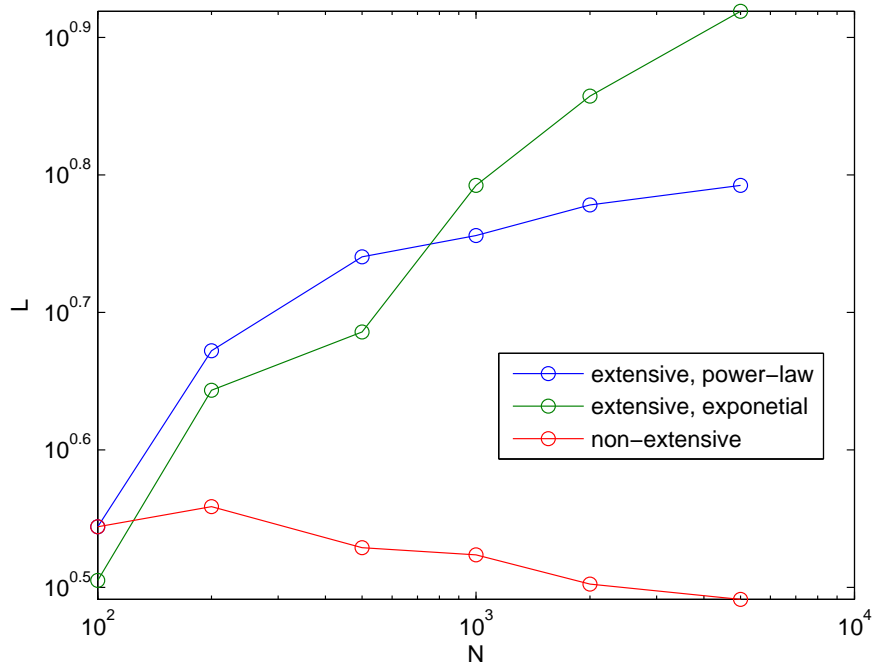} \\
\label{pathlength}
\end{center}
\end{figure}

 According to Fig.\ref{pathlength}, the RIPA under non-extensive limit is always a small world. The average path-length even lightly decays as $N$ grows. This observation can be interpreted as the transportation in a fixed area becomes more convenient when you have more choices of transition points.
We also observe that the RIPA under extensive limit is a small world when the relevance function $f$ has the power-law decay ($f(d)=d^{-2}$), but is not when $f$ has a exponential decay ($f(d)=e^{-\lambda d}$). From the physics aspect, the two relevance functions are analogues of long-range and short-range correlations. So this observation can be concluded as the RIPA network is a small world when the relevance function represents a long-range correlation.

The following theorem give a criterion when the RIPA network on two-dimensional space is not a small world.\\
{\bf Theorem}: The network is not a small world network if the
\begin{equation}
\lim_{a\rightarrow \infty} a^2\int_{L=a}^{\infty} Lf(L)dL =0.
\label{criterion}
\end{equation}
\textbf{Proof:} First, we show that the probability distribution $p(L)$ of $L$, the length of the links, is proportional to $Lf(L)$. For a fixed vertex $i$ at the location $x_i$ at arbitrary time step, consider the length of the next link attached to it. Ignoring the boundary effect of the two dimensional space (for unit square it means $L<1/2$), the probability that the new vertex $j$ appears at $x_j$ which is apart from $x_i$ with the distance $L$ and attaches to the vertex $i$ is
$$\oint_{R(x_i,L)} {z(x_j)\over Z} {k_i \rho(x_i,x_j)\over z(x_j)} dx_j=\oint_{R(x_i,L)}{k_i f(L)\over Z} dx_j\sim Lf(L) $$
where $R(x_i,L)$ is the circle centered at $x_i$ with radius $L$. Since for all of the preexisting vertices, the distribution of the length of the next new link is the same, so is the overall length distribution of the next new link at arbitrary time step. So except for the $m_0$ initial links which can be neglected in the large N limit, the length distribution $p(L)$ is proportional to $Lf(L)$.

Then we divide the two-dimensional space into blocks with edge length $a$. In the extensive large $N$ limit, the density of vertices $\rho_0$ keeps constant, so the expected number of links which is attached to the given block and longer than $a$ is $\rho_0 a^2\int_{L=a}^{\infty} Lf(L)dL$. If Eqn. (\ref{criterion}) holds, for big enough $a$, the probability to find a link longer than $a$ in a given block can be controlled by arbitrarily small $\epsilon>0$, i.e. with probability $1-\epsilon$ one can only move to its neighboring blocks by one step along the path. Therefore, the shortest path length between two vertices with distance $D$ is lower bounded by ${D\over a}(1-\epsilon)^{D/a}$ which obviously is not a small world. Similar criterion is easy to establish for $R^n$ space.


\section{core-periphery structure}
\begin{figure}[!htbp]
\caption{Giant cluster size $g$ after removal $f_r$ fraction of nodes in a descending order of degree.}
\begin{center}
\includegraphics[width=0.9\textwidth]{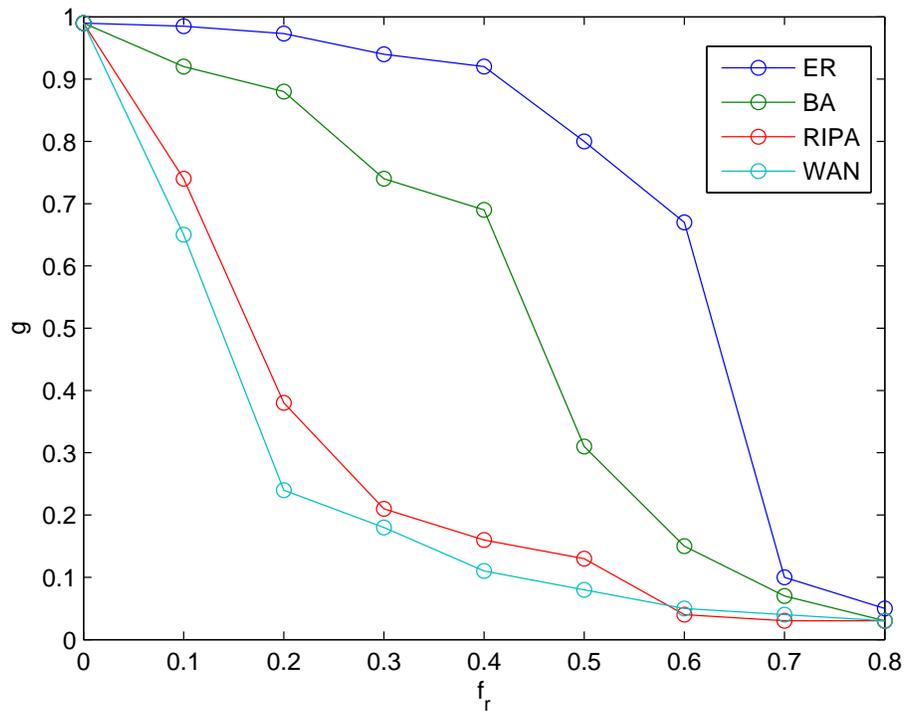} \\
\label{robust}
\end{center}
\end{figure}

Core-periphery structure is observed in several real world complex networks\cite{Csermely,ROMBACH}. In the network with such kind of structure, there is a subnetwork called ``core'' which is tightly connected, and the complementary subnetwork, the periphery, are fragmental and mostly attached to the core. A significant feature of the core-periphery structure is that the network is vulnerable to the attacks on the core\cite{ROMBACH}. By successively removing nodes from the core, the whole network will quickly fall into several disconnected parts. The Fig.\ref{robust} shows how the giant cluster size decreases as the nodes are removed in the descending order of the degrees. As shown in the figure, the BA network has hubs therefore are more vulnerable to the attacks on the high degree nodes than the ER networks, but it still has a high threshold (about $0.5$ in the figure) when the giant cluster size has a fast decay. For RIPA and WAN, however, the giant cluster sizes both decrease quickly at the very beginning. So the RIPA network model captures the core-periphery structure as in the WAN network.

\subsection{RIPA on the Sphere}
\begin{figure}[!htbp]
\caption{Network generated on sphere with $m=3$, $N=5000$, $\lambda=5$. Two plots are the views of the same sphere from different angles. The color(brightness) indicates the logarithm of local partition.}
\begin{center}
\includegraphics[width=0.4\textwidth,bb=168 256 460 550]{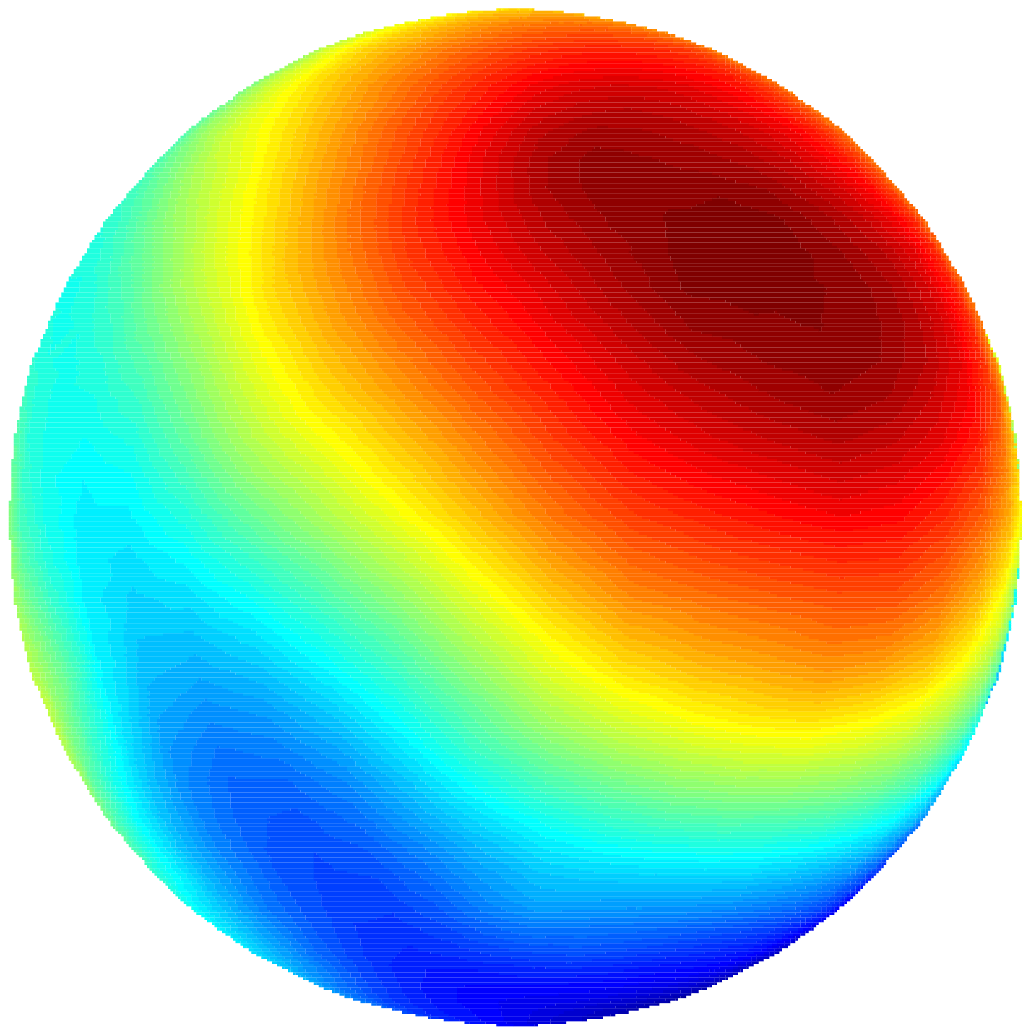}\\
\includegraphics[width=0.4\textwidth,bb=168 256 460 550]{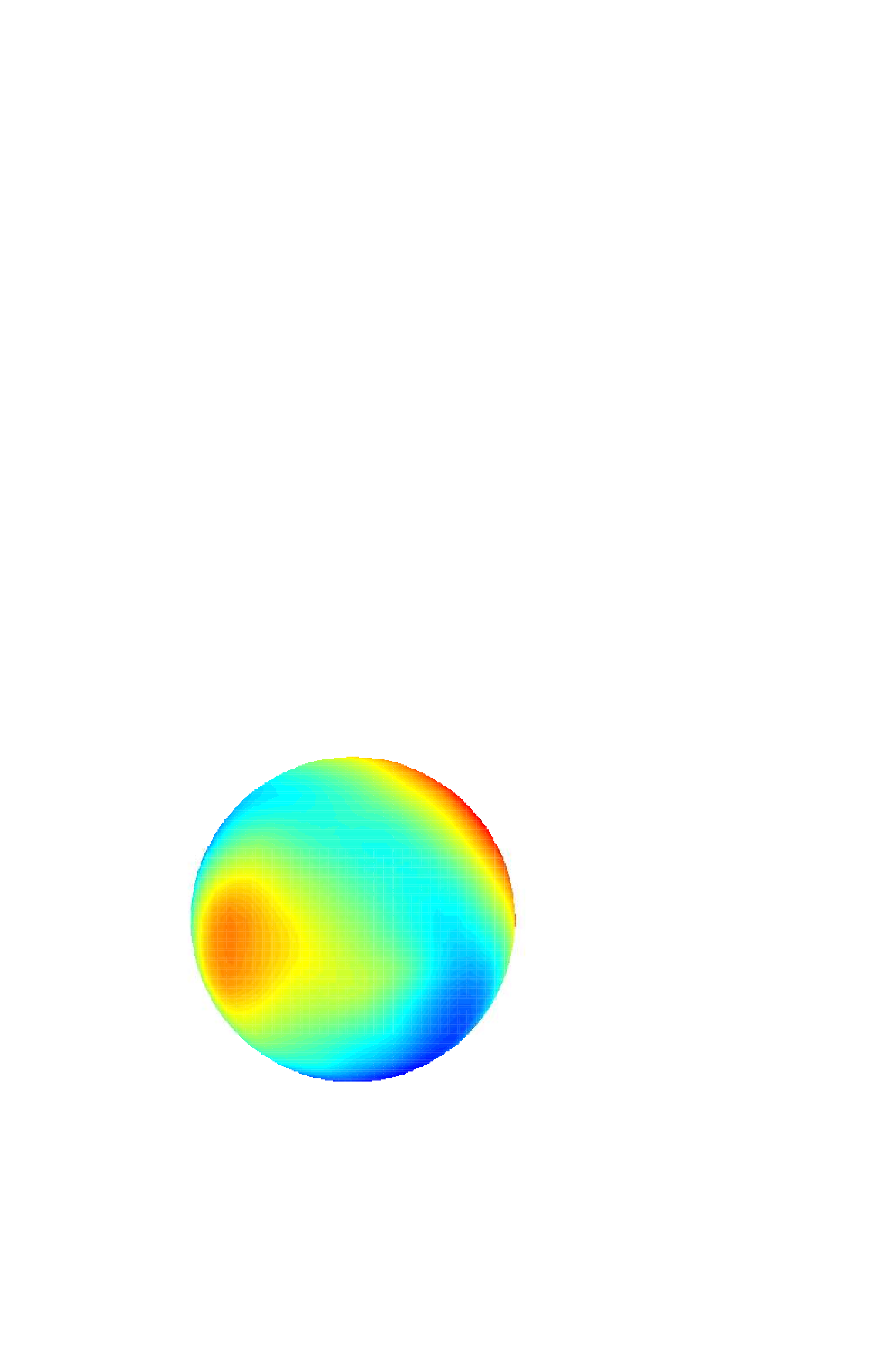}\\
\label{sphere}
\end{center}
\end{figure}

Similarly, we implement the RIPA on the sphere where the metric is given by spherical distance. As shown in Fig.\ref{sphere}, the . Interestingly, some qualitative behavior is quite stable in the simulations, eg. the spherical angle between the first two largest hubs are usually around $0.6\pi-0.7 \pi$. However, this network is still quite far from the case of the earth. On the earth, city can only locate on the continents, and the metric is not uniform. The oceans, rivers and mountains may affect the effective distance.

\section{Discussion}
In this paper we investigate a very simple case, RIPA on two dimensional space, corresponding to the scenario of the between-city transportation. WAN fits this case best, since for airlines the relevance of two cities are mainly decided by the geographic distance. But obviously there are other factors affect the relevance of two cities, e.g. the cities belonging to two different countries should have weaker relevance than those of the same country.

For other cases, the underlying metric space of relevance may be even more complicated and hard to describe. For example, in the social networks, the relevance between individuals may involve the geographic distance, the cultural difference, the social stratification, the political stands and so on. Further, this metric can be changed by the development of technologies in transportation and communication. However, previous studies of social networks have shown some evidence for the existence of the underlying metric space\cite{Milgram,Kleinberg}. In conclusion, the underlying metric space of the relevance is the main source of the network complexity. Although it is usually hard to give a full description of the metric space in the real world cases, analyzing different metric space and their corresponding RIPA networks will help us to understand the global structure of real world networks.

\section*{Acknowledgment}

This work was supported in part by the Army Research Laboratory under Cooperative Agreement Number W911NF-09-2-0053, by the Army
Research Office Grant Nos  W911NF-09-1-0254 and W911NF-12-1-0546, and by the Office of Naval Research Grant No. N00014-09-1-0607.
The views and conclusions contained in this document are those of the authors and should not be interpreted as representing the
official policies either expressed or implied of the Army Research Laboratory or the U.S. Government.


%

\newpage

\bibliographystyle{}

\end{document}